\documentclass[aps,10pt,prl,twocolumn,floatfix,nobibnotes,superscriptaddress]{revtex4-2}
\usepackage{dcolumn,amsmath}
\usepackage{bm}
\usepackage{hyperref}
\usepackage{xcolor}             
\usepackage[utf8]{inputenc}
\usepackage{graphicx}
\usepackage{dcolumn}
\usepackage{bm}
\usepackage{amsmath}
\usepackage{color} 
\usepackage{soul}
\usepackage[normalem]{ulem}
\usepackage{hyperref}
\usepackage[english]{babel}
\usepackage{physics}
\usepackage{tabularx}

\setcitestyle{super}
\usepackage{etoolbox}
\patchcmd{\section}
  {\centering}
  {\raggedright}
  {}
  {}
\patchcmd{\subsection}
  {\centering}
  {\raggedright}
  {}
  {}
\usepackage[normalem]{ulem}

\makeatletter
\renewcommand{\fnum@figure}{Fig.~\thefigure}
\renewcommand{\fnum@table}{Table~\thetable}
\def\@hangfrom@section#1#2#3{\@hangfrom{#1#2}#3}
\def\@hangfroms@section#1#2{#1#2}
\renewcommand\section{\@startsection {section}{1}{0pt}
  {-3.5ex \@plus -1ex \@minus -.2ex}
  {2.3ex \@plus.2ex}
  {\normalfont\large\bfseries}}
\makeatother

\begin{document}

\author{Qi Xiao}
\thanks{These authors contributed equally to this work.}
\affiliation{State Key Laboratory of Low-Dimensional Quantum Physics, Department of Physics, Tsinghua University, Beijing 100084, China}

\author{Gleb Penyazkov}
\thanks{These authors contributed equally to this work.}
\affiliation{State Key Laboratory of Low-Dimensional Quantum Physics, Department of Physics, Tsinghua University, Beijing 100084, China}

\author{Xiangliang Li}
\thanks{These authors contributed equally to this work.}
\affiliation{Beijing Academy of Quantum Information Sciences, Beijing 100193, China}

\author{Beichen Huang}
\affiliation{State Key Laboratory of Low-Dimensional Quantum Physics, Department of Physics, Tsinghua University, Beijing 100084, China}

\author{Wenhao Bu}
\affiliation{Beijing Academy of Quantum Information Sciences, Beijing 100193, China}

\author{Juanlang Shi}
\affiliation{State Key Laboratory of Low-Dimensional Quantum Physics, Department of Physics, Tsinghua University, Beijing 100084, China}

\author{Haoyu Shi}
\affiliation{State Key Laboratory of Low-Dimensional Quantum Physics, Department of Physics, Tsinghua University, Beijing 100084, China}

\author{Tangyin Liao}
\affiliation{Department of Precision Instrument, Tsinghua University, Beijing 100084, China}

\author{Gaowei Yan}
\affiliation{State Key Laboratory of Low-Dimensional Quantum Physics, Department of Physics, Tsinghua University, Beijing 100084, China}

\author{Haochen Tian}
\affiliation{Division of Time and Frequency Metrology, National Institute of Metrology, Beijing 100029, China
}

\author{Yixuan Li}
\affiliation{State Key Laboratory of Low-Dimensional Quantum Physics, Department of Physics, Tsinghua University, Beijing 100084, China}

\author{Jiatong Li}
\affiliation{State Key Laboratory of Low-Dimensional Quantum Physics, Department of Physics, Tsinghua University, Beijing 100084, China}

\author{Bingkun Lu}
\affiliation{Division of Time and Frequency Metrology, National Institute of Metrology, Beijing 100029, China
}

\author{Li You}
\affiliation{State Key Laboratory of Low-Dimensional Quantum Physics, Department of Physics, Tsinghua University, Beijing 100084, China}
\affiliation{Beijing Academy of Quantum Information Sciences, Beijing 100193, China}
\affiliation{Frontier Science Center for Quantum Information, Beijing 100084, China}

\author{Yige Lin}
\affiliation{Division of Time and Frequency Metrology, National Institute of Metrology, Beijing 100029, China
}

\author{Yuxiang Mo}
\affiliation{State Key Laboratory of Low-Dimensional Quantum Physics, Department of Physics, Tsinghua University, Beijing 100084, China}

\author{Shiqian Ding}
\thanks{Corresponding author: dingshq@mail.tsinghua.edu.cn}
\affiliation{State Key Laboratory of Low-Dimensional Quantum Physics, Department of Physics, Tsinghua University, Beijing 100084, China}
\affiliation{Beijing Academy of Quantum Information Sciences, Beijing 100193, China}
\affiliation{Frontier Science Center for Quantum Information, Beijing 100084, China}

\title{A continuous-wave vacuum ultraviolet laser for the nuclear clock}

\begin{abstract}

\textbf{The exceptionally low-energy isomeric transition in $^{229}$Th at around 148.4 nm offers a unique opportunity for coherent nuclear control and the realisation of a nuclear clock. Recent advances, most notably the incorporation of large ensembles of $^{229}$Th nuclei in transparent crystals and the development of pulsed vacuum‑ultraviolet (VUV) lasers, have enabled initial laser spectroscopy of this transition. However, the lack of an intense, narrow-linewidth VUV laser has precluded coherent nuclear manipulation. 
Here we introduce and demonstrate the first continuous-wave laser at 148.4 nm, generated via four-wave mixing (FWM) in cadmium vapor. The source delivers 100 nW of power with a linewidth well below 100 Hz and supports broad wavelength tunability. This represents a five-orders-of-magnitude improvement in linewidth over all previous single-frequency lasers below 190 nm, marking a major advance in laser technology.
We develop a spatially resolved homodyne technique to place a stringent upper bound on the phase noise induced by the FWM process and demonstrate sub-hertz linewidth capability. 
These results eliminate the final technical hurdle to a $^{229}$Th‑based nuclear clock, opening new directions in quantum metrology, nuclear quantum optics and precision tests of the Standard Model. More broadly, they establish a widely tunable, ultranarrow-linewidth laser platform for applications across quantum information science, condensed matter physics, and high-resolution VUV spectroscopy. 
}
\end{abstract}

\flushbottom
\maketitle
The coherent excitation of nuclear transitions has long remained elusive due to the vast energy mismatch between nuclear states and available laser sources. A striking exception is the uniquely low-energy isomeric transition in $^{229}$Th at 8.4 eV ($\sim$ 148.4 nm)~\cite{1976Energy,2007Energy,2019EnergyMunich,2019EnergyJapan,2020Energy,2022EnergyCERN,LaserSpecSchumm,LaserSpecHudson,LaserSpecYe,ThoriumReviewPeik}, which combines optical accessibility, as finally enabled in this work, with extraordinary insensitivity to external perturbations, owing to the small nuclear electric and magnetic moments. These properties provide a foundation for a revolutionary nuclear clock~\cite{ClockPeik} with unprecedented precision~\cite{ClockKuzmich},
and offer a powerful probe for testing physics beyond the Standard Model~\cite{FlambaumVariation,CrystalHudson,DarkMatter2025}.

Recent breakthroughs using $^{229}$Th-doped crystals~\cite{ClockPeik,CrystalHudson,CrystalSchumm,2022EnergyCERN,CrystalSchummExp} have enabled laser spectroscopy of the isomer~\cite{LaserSpecSchumm,LaserSpecHudson,LaserSpecYe,ThF4,Temperature,QuenchingHudson,QuenchingPeik} using pulsed vacuum‑ultraviolet (VUV) sources, including nanosecond four-wave mixing~\cite{HudsonVUV,PeikVUV} and high-harmonic generation frequency combs~\cite{ChuankunVUV}. 
These systems allow simultaneous interrogation of up to $10^{15}$ nuclei, enabling fluorescence detection even with weak laser fields that excite only a small fraction of the ensemble.
However, existing pulsed lasers suffer from either broad linewidths (GHz-scale)~\cite{HudsonVUV,PeikVUV} or extreme power dilution across millions of comb teeth, with only 1 nW residing in the desired tooth~\cite{ChuankunVUV}.
As a result, the achievable power spectral density remains far too low for coherent control.
This limitation is even more acute in ion-trap platforms~\cite{ClockPeik,ClockKuzmich,KuzmichCrystal,Th2IsomerSpec,IonThirolf,IonPeik,Th3JapanNature}, where only a small number of $^{229}$Th ions are available. While these systems offer exceptional isolation, ultralong interrogation times, and the full toolbox of quantum manipulation~\cite{Leibfried2003RMP,QLSAl}, including optical nuclear-qubit encoding and entanglement protocols, laser excitation of the nucleus has yet to be demonstrated. The lack of sufficient laser power spectral density remains the central bottleneck impeding progress toward ultimate nuclear clock performance~\cite{ClockKuzmich} and nuclear-based quantum information processing. 

\begin{figure*}[htbp]
 \includegraphics[width=179mm]{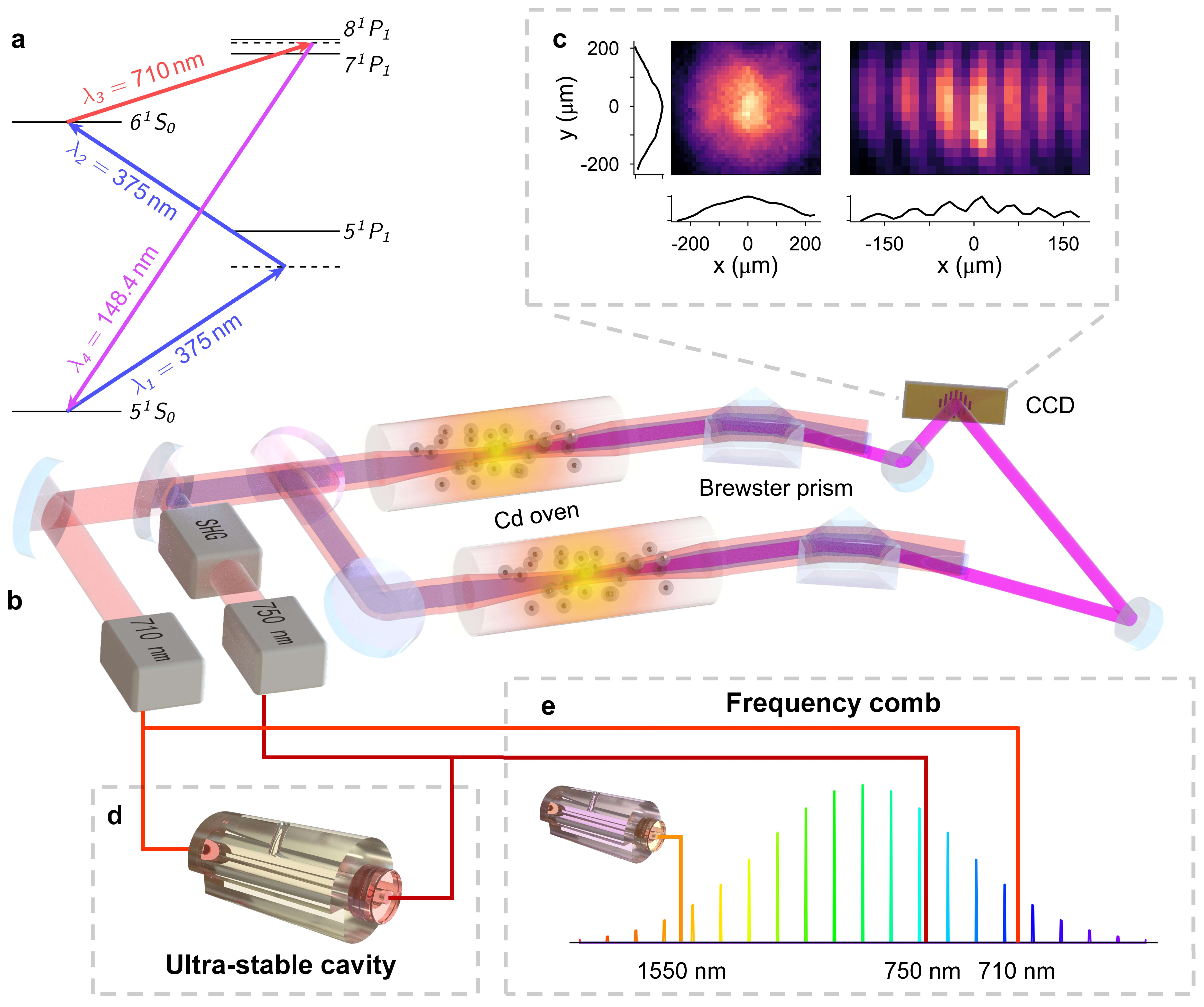}
   \caption{\textbf{Generation of a CW VUV laser with narrow linewidth.}
  \textbf{a}, Energy-level diagram for the resonance-enhanced FWM process in cadmium vapor.
  \textbf{b}, Schematic of the experimental setup.
  The 375 nm and 710 nm beams are combined and focused into two cadmium ovens. The resulting VUV beams are separated from the fundamental beams using Brewster-angle MgF$_2$ prisms and interfered on a CCD camera. 
  SHG: second-harmonic generation.
   \textbf{c}, Spatial mode profile of the generated 148.4 nm beam (left panel) and interference pattern (right panel) recorded with a 10 s exposure time. Black curves show 1D projections along the horizontal and vertical axes.
   \textbf{d}, Both Ti:sapphire lasers are frequency-stabilised to a shared ultrastable ULE cavity to compress their linewidths.
   \textbf{e}, Linewidths of the 750~nm and 710~nm lasers are probed via beat notes with an optical frequency comb referenced to a ULE-cavity-stabilised 1550 nm laser.}
    \centering
    \label{fig1}
\end{figure*}

Despite its central importance, continuous-wave (CW) laser radiation at 148.4 nm, concentrating all power into a narrow linewidth to boost power spectral density over pulsed sources, has remained a formidable challenge~\cite{ThoriumReviewPeik}. 
Narrow-linewidth CW VUV sources are also in demand across other research areas, including high-resolution VUV spectroscopy, angle-resolved photoemission spectroscopy (ARPES)~\cite{ARPES} in condensed matter systems, and direct laser cooling of Al$^+$ ions for state-of-the-art optical clocks~\cite{QLSAl,AlClock}. 
However, standard phase-matched frequency conversion in this spectral region is precluded by strong absorption and dispersion in nonlinear crystals~\cite{DeapUV}, 
while quasi-phase matching in periodically poled VUV-transparent nonlinear crystals demands domain periods of microns or less~\cite{QPM2009} that remain beyond current fabrication capabilities~\cite{QPMPeriod}. 
Four-wave mixing (FWM) in atomic vapors~\cite{TwoResonance,Hansch1999, CW2012} offers an alternative approach, exploiting third-order nonlinearities to generate sum-frequency radiation. This method has enabled CW generation of the hydrogen Lyman-$\alpha$ line in mercury~\cite{Hansch1999}, but with estimated linewidths at 10 MHz~\cite{Hansch2005}, which are far too broad for nuclear clock applications.

Here we report, to our knowledge, the first CW laser at the $^{229}$Th isomeric transition wavelength~\cite{arxivpaper}, realised via resonance-enhanced FWM in cadmium vapor. 
This laser delivers 100 nW of CW power at 148.4 nm with a linewidth well below 100 Hz, and offers continuous tunability across at least 140 nm to 175 nm using cadmium~\cite{arxivpaper}, with broader coverage feasible using other nonlinear media. 
This represents several orders of magnitude improvement in power spectral density over existing VUV sources~\cite{HudsonVUV,ChuankunVUV,PeikVUV} below 190~nm~\cite{CW191nm}, providing the critical enabling tool for coherent manipulation of the $^{229}$Th nucleus. 
We further show that the FWM process preserves sub-hertz coherence, with an upper-bound fractional frequency instability well below $10^{-16}$. This resolves key concerns about whether GHz-scale Doppler and collisional broadening in hot vapor media would degrade phase coherence in CW nonlinear frequency conversion~\cite{VUVCoherence2014}. The phase noise is measured using a novel spatially resolved homodyne technique that operates with only sub-nanowatt optical power, which may become broadly useful for measuring optical phase in power-limited regimes.
Finally, our measured power agrees well with predictions from a theoretical model~\cite{arxivpaper} that combines FWM theory with \textit{ab initio} atomic structure calculations~\cite{arxivpaper2}, providing predictive power for future source design without relying on extensive experimental iteration.
Together, these results establish a robust and versatile CW laser platform with ultranarrow linewidth and wide tunability in the VUV, opening new frontiers in nuclear quantum optics, quantum metrology, condensed matter physics, quantum information processing, and chemical dynamics.

\vspace{0.6cm}

\noindent \textbf{VUV generation protocol}

We generate the 148.4 nm radiation via resonance-enhanced FWM in cadmium vapor.
Two photons ($\lambda_1$ and $\lambda_2$) at 375 nm, tuned to a two-photon resonance with the $6\: ^1S_0$ state, and a third photon ($\lambda_3$) at 710 nm are combined and focused into a 50 cm-long cadmium oven held near $T = 550\,^\circ\mathrm{C}$, with a shared confocal parameter of $b\sim5$ mm (see Fig.~\ref{fig1}). The fundamental beams originate from two Ti:sapphire lasers, each locked to a common ultrastable ultralow expansion (ULE) cavity to ensure narrow linewidths. The generated VUV beam is spatially separated from the fundamental beams using a Brewster-angle MgF$_2$ prism for 148.4 nm, minimizing reflective losses at both input and output facets. The VUV signal is detected using a solar-blind photomultiplier tube (PMT).

To evaluate phase coherence, the fundamental beams are split and directed into two independent cadmium ovens. 
The resulting VUV beams are overlapped on a CCD camera with a small crossing angle, producing high-contrast interference fringes that allow direct extraction of the relative optical phase (see Methods).

\vspace{0.6cm}
\noindent \textbf{VUV yield characterisation}

\begin{figure*}
\centering
    \includegraphics[width = 179mm]{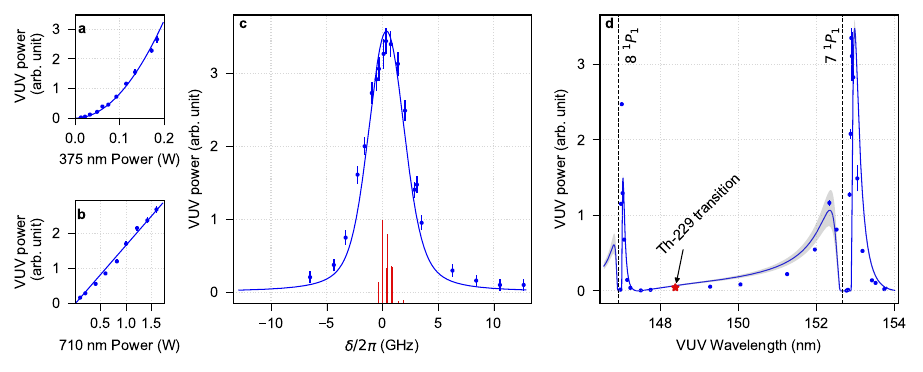}
    \caption{\textbf{VUV yield characterisation.}
    VUV yield as a function of \textbf{a}, 375 nm laser power and \textbf{b}, 710 nm laser power. Solid lines are quadratic and linear fits, respectively. 
    \textbf{c}, Two-photon resonance profile of the $6\: ^1S_0$ state in naturally abundant cadmium. The horizontal axis represents detuning from the $^{114}$Cd isotope.
    Blue points show experimental data; the solid blue curve is the scaled theoretical prediction at $T = 525\,^\circ$C and 70 mbar argon pressure.
    Red vertical lines mark isotope shifts and natural abundances. 
    \textbf{d}, VUV output versus wavelength ($\lambda_4$), corresponding to variation in the third-photon wavelength ($\lambda_3$). The solid blue line is the theoretical prediction (scaled by 0.35) at $b = 4.1~\mathrm{mm}$ and $T = 525\,^\circ\mathrm{C}$. 
    The shaded band reflects uncertainty in the confocal parameter $b = 4.1\pm 0.2 ~\text{mm}$.
    All error bars represent 1$\sigma$ statistical uncertainties.}
    \centering
    \label{VUVoutput}
\end{figure*}

The generated VUV power, $P_4$, scales as $P_{4}\propto  P_{1,2} ^2 P_{3} \times |\chi_a^{(3)}|^2 \times G$, where $P_{1,2}$ and $P_{3}$ denote the input powers at 375~nm and 710~nm, respectively, $\chi_a^{(3)}$ is the third-order nonlinear susceptibility per cadmium atom, and $G$ is the phase-matching factor~\cite{arxivpaper} governed by dispersion in the cadmium vapor and argon buffer gas, as well as the Gouy phase shift due to tight focusing (see Methods). 
As shown in Fig.~\ref{VUVoutput}a and Fig.~\ref{VUVoutput}b, the measured VUV power exhibits a quadratic dependence on $P_{1,2}$ and a linear dependence on $P_{3}$, consistent with the FWM scheme involving two photons at 375~nm and one at 710~nm.
No saturation is observed across the explored power range, suggesting additional headroom for increased VUV yield.

Near the two-photon resonance condition, cadmium atoms exhibit enhanced nonlinear susceptibility $\chi^{(3)}$~\cite{VidalReview,arxivpaper}. 
The resonance profile is shaped by Doppler and pressure broadening, as well as isotope shifts of the $6\: ^1 S_0$ state in naturally abundant cadmium vapor.
We characterise this profile by scanning the 375 nm laser frequency while monitoring the VUV output (Fig. \ref{VUVoutput}c). The observed full width at half maximum (FWHM) of $4.4(2)$~GHz is in reasonable agreement with the theoretical prediction of 3.8~GHz~\cite{arxivpaper}. 
Notably, this width exceeds the isotope shifts, confirming that all cadmium isotopes contribute coherently to the FWM process. As a result, naturally abundant cadmium produces VUV output comparable to isotopically enriched samples~\cite{arxivpaper}.

To explore the tunability of our source, we vary the wavelength of the third photon ($\lambda_3$) while recording the VUV power (Fig.~\ref{VUVoutput}d). Tuning $\lambda_3$ from 678.9~nm to 851.8~nm corresponds to VUV output spanning 146.97~nm to 153.7~nm. 
A pronounced enhancement up to 79-fold is observed at 152.9~nm relative to 148.4~nm. This is attributed to third-photon resonance enhancement of $\chi^{(3)}$ near the $7\: ^1 P_1$ state~\cite{arxivpaper} (see Methods). A similar feature appears near the $8\: ^1 P_1$ state at 147.01~nm. 
Additionally, our calculations~\cite{arxivpaper,arxivpaper2} predict a vanishing $\chi^{(3)}$ around 147.5~nm due to destructive interference among contributing intermediate states, leading to a complete suppression of VUV generation, again consistent with experiment.

By coupling the fundamental beams into optical fibers for spatial mode filtering and minimizing aberrations along the propagation path, we reach a 148.4~nm output power that is approximately half the theoretical prediction. 
This close agreement validates our model~\cite{arxivpaper,arxivpaper2}. Remaining discrepancy are likely due to measurement uncertainty in 148.4~nm power, spatial inhomogeneity in the vapor, residual wavefront distortions, and limited precision in atomic transition parameters.

\begin{figure}
\includegraphics{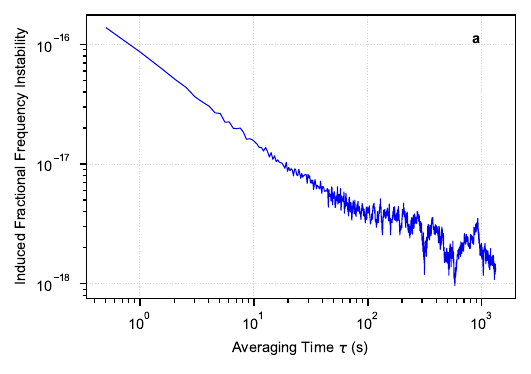}
\includegraphics{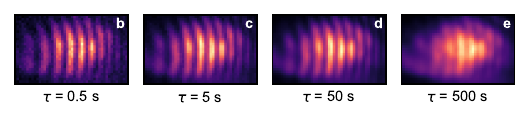}
   \caption{\textbf{Upper bound on phase noise induced by the FWM process.} \textbf{a}, Allan deviation of the induced fractional frequency instability for a single VUV beam, extracted from a time series of interference patterns. The value is scaled by $\sqrt2$ under the assumption that the two VUV beams contribute uncorrelated noise of equal magnitude. \textbf{b-e}, 
   Representative interference patterns used for the Allan deviation analysis, showing phase stability over different averaging times $\tau$. Each pattern is constructed by integrating multiple 0.1 s exposure images. The total averaging time $\tau$ includes a 0.4 s dead time between exposures.
   }
    \centering
    \label{beating_analysis}
\end{figure} 

\vspace{0.6cm}
\noindent \textbf{Phase coherence of the VUV beam}

We demonstrate that the phase noise introduced by the FWM process is sufficiently low for nuclear clock applications. This is verified by interfering two VUV beams generated in independent cadmium ovens and recording high-contrast fringes on a CCD camera with a 10 s exposure time (see Fig.~\ref{fig1}c). The preserved fringe visibility over this integration time proves excellent phase coherence. 
Our measurements also indicate that the VUV linewidth is not broadened by Doppler effects in the hot cadmium vapor~\cite{VUVCoherence2014}. This insensitivity arises from a precise cancellation: the Doppler shifts of the fundamental beams in the atomic rest frame are exactly offset by the frequency transformation of the generated VUV field back into the laboratory frame.
Moreover, the absence of observable broadening confirms that nonlinear optical processes remain robust against high collision rates in CW operation, consistent with their reliance on virtual intermediate states that are immune to collisional dephasing.

To quantify low-frequency phase stability, we extract the relative phase from fringe displacement (see Methods) across a sequence of 10,000 interference images, each with 0.1 s exposure. The resulting Allan deviation of the extracted phase reveals an induced single-beam fractional frequency instability of $8.6\times 10^{-17}$ at 1 s averaging time (see Fig.~\ref{beating_analysis}a), assuming equal and uncorrelated noise contributions from the two beams. The observed scaling with averaging time suggests that white frequency noise dominates.
To probe high-frequency phase noise beyond the resolution bandwidth of the fringe displacement method, we analyze the fringe visibility for 1 s exposures (see Methods). 
Assuming a Lorentzian spectral profile, the measured visibility corresponds to an FWHM linewidth of 0.08(2) Hz, implying that approximately 97\% of the optical power is confined within a 1 Hz bandwidth.

These measurements should be interpreted as upper bounds on the phase noise introduced by the FWM process, since residual technical noise such as mechanical vibrations and air turbulence has not been excluded.
Notably, our spatially resolved homodyne technique leverages spatial multiplexing to reject DC and low-frequency noise, enabling robust and sensitive phase extraction at sub-nanowatt optical power levels, well below the sensitivity threshold of conventional homodyne detection.

Since both VUV beams originate from the same fundamental lasers, these phase noise measurements are immune to common-mode noise of the inputs.
With the FWM process contributing negligible additional phase noise, the ultimate VUV linewidth is primarily limited by the stability of the fundamental lasers.
In our setup, both the 750 nm and 710 nm Ti:sapphire lasers are locked to a ULE cavity and exhibit linewidths near 2 Hz. 
Under the most conservative assumption of fully correlated noise between the fundamental beams, the resulting VUV linewidth is expected to be 50 Hz, based on scaling relations under frequency multiplication~\cite{VUVCoherence2014,PeikVUV} (see Methods).

The VUV linewidth can be further reduced to the sub-10 Hz level by improving the fundamental laser stabilisation.
For 100~nW of power focused to a $2$~$\mu$m diameter spot, the criterion for observing nuclear Rabi oscillations~\cite{LarsCalculation,arxivpaper}, $\Gamma_l < 2\Omega_{eg} \approx 2\pi \times 91\;\text{Hz}$, is readily fulfilled, where $\Gamma_l/2\pi$ and $\Omega_{eg}$ denote the VUV laser linewidth and the nuclear Rabi frequency, respectively. 

The CW nature of our source eliminates the broad spectral background inherent
to pulsed VUV sources~\cite{HudsonVUV,ChuankunVUV,PeikVUV}, thereby reducing optical damage to $^{229}$Th-doped crystals~\cite{LaserSpecSchumm} and suppressing uncontrolled effects such as optical quenching~\cite{QuenchingJapan,QuenchingPeik,QuenchingHudson} and light-induced frequency shifts that degrade measurement precision. 
Finally, the demonstrated phase coherence in the FWM process allows the VUV phase to be transferred to the fundamental lasers, which can then be locked to a standard optical frequency comb, bypassing the need for the challenging VUV comb at 148.4~nm~\cite{ChuankunVUV}. This establishes a coherent and traceable phase link between the VUV field and the microwave domain for clock operation.

\vspace{0.6cm}
\noindent \textbf{Conclusion}

We have demonstrated a continuous-wave laser at 148.4 nm with narrow linewidth and sufficient intensity for coherent control of the $^{229}$Th nuclear isomer transition.
Using a spatially resolved homodyne technique, we place stringent upper bounds on the phase noise introduced by four-wave mixing in hot vapor, verifying sub-hertz coherence.
This overcomes the final technical barrier to a $^{229}$Th-based nuclear clock and establishes a robust pathway toward coherent optical manipulation of nuclear states. 
The generated vacuum ultraviolet power could be further enhanced, potentially beyond 100~$\mu$W, with higher fundamental beam powers, optical enhancement cavities, or alternative nonlinear media with third-photon resonance enhancement near 148.4~nm.

Beyond enabling the $^{229}$Th nuclear clock and nuclear quantum optics, the laser system reported here, featuring excellent coherence, continuous-wave operation, and broad wavelength tunability, opens new directions across a wide range of disciplines.
A natural extension of the platform enables generation of intense 167.1 nm laser, which, by fortunate coincidence, benefits from strong third-photon resonance in cadmium. This wavelength directly addresses the famous long-standing challenge of laser cooling and state detection of Al$^+$ ions, eliminating the need for quantum logic spectroscopy and significantly simplifying and improving Al$^+$ optical clocks~\cite{QLSAl}, which currently define the state of the art in clock accuracy~\cite{AlClock}.
In quantum information science, our source could catalyze progress in the emerging Rydberg-ion platform for universal quantum computing~\cite{RydbergIon,GateRydbergIon}. 
While two-qubit gates have been demonstrated using two-photon excitation to Rydberg states~\cite{GateRydbergIon}, direct VUV excitation, originally proposed but previously inaccessible, would eliminate intermediate-state decoherence and enable faster, higher-fidelity operations. 
In condensed matter physics, the unprecedented spectral purity of our source could dramatically enhance the energy resolution of ARPES, potentially revealing subtle many-body correlations and fragile quasiparticle states in topological materials and high-$T_c$ superconductors~\cite{ARPES}. 
In the chemical sciences, the system could enable ultrahigh-resolution photoelectron spectroscopy (PES) of molecules, clusters, and transient intermediates, providing new insights into electronic structure and reaction dynamics~\cite{Chemistry2}.
Together, these capabilities establish our VUV laser system as a broadly enabling tool, with transformative potential across quantum science, precision metrology, condensed matter physics, and chemical dynamics.

\noindent \textbf{Acknowledgments.}
We thank X. Zhang and S. Zhou for discussions, Z. Xu and X. Wen for advice on frequency doubling, M. K. Tey for support on electronics, and C. Zhang, H. Wu and J. Ye for comments on the manuscript. 
This work is supported by the National Natural Science Foundation of China (No. 12341401, No. 12274253, and No. 92265205) and the Shanghai Municipal Science and Technology Commission (No. 25LZ2600402).

\noindent \textbf{Author contributions.}
Q.X., G.P., X.L., B.H., W.B., J.S., H.S., G.Y., Y.Li, J.L., L.Y., Y.M. and S.D. designed the experiment, constructed the setup, and carried out the measurements; T.L., H.T., B.L. and Y.Lin built the ULE-cavity-stabilised 1550 nm laser; Q.X. and S.D. wrote the manuscript with input from all authors. 

\noindent \textbf{Competing interests.}
The authors declare no competing interests.

\noindent \textbf{Correspondence and requests for materials} should be addressed to Shiqian Ding.

\makeatletter
\def\endthebibliography{%
    \def\@noitemerr{\@latex@warning{Empty `thebibliography' environment}}%
    \endlist
}
\makeatother

\section{Methods}

\noindent \textbf{VUV generation and diagnostics}

Cadmium vapor is selected as the nonlinear medium for its large third-order susceptibility and compatibility with high-power laser wavelengths at 375~nm and 710~ nm~\cite{arxivpaper,arxivpaper2}. 
The two 375 nm photons couple the $5\: ^1S_0$ and $6\: ^1S_0$ states, while the 710 nm photon completes the sum-frequency process to generate 148.4 nm light.
The horizontally polarised fundamental beams, 2~W at 375~nm and 4~W at 710~nm, are provided by two Ti:sapphire lasers. The 375 nm light is produced via second-harmonic generation in a bow-tie enhancement cavity containing a lithium triborate (LBO) crystal. 
Both Ti:sapphire lasers are locked to a common 10 cm-long ULE cavity with finesse $\sim 2\times10^5$. Their linewidths are characterised using a frequency comb referenced to a 1 Hz-linewidth ULE-stabilised erbium-doped fiber laser at 1550 nm. The fundamental beams are focused into a 50 cm-long cadmium oven with a shared confocal parameter of $b \sim 5$ mm. The oven has a $\sim 5$ cm central hot zone, operated near 550~°C, with 70 mbar of argon buffer gas to inhibit cadmium condensation on the windows. The input and output windows are AR-coated fused silica and uncoated MgF$_2$, respectively, and are maintained near room temperature with water cooling.

The generated VUV beam is spatially separated using a 68$^\circ$ MgF$_2$ prism, which satisfies the Brewster condition for both input and output beams to minimize reflective loss. The vacuum chamber interior is blackened with carbon nanotube coating to suppress stray light. VUV output is measured using a solar-blind PMT, shielded by a 30 dB neutral density filter. Accounting for window transmission, MgF$_2$ lenses losses, and PMT quantum efficiency, the overall photon detection efficiency is estimated to be $1 \times 10^{-4}$.
To assess coherence, the fundamental beams are split and sent into two identical cadmium ovens. The resulting VUV beams interfere at a 0.2$^\circ$ angle on a front-end-open CCD camera, with two narrow-band filters (30\% transmission at 148.4 nm) suppressing background.

\noindent \textbf{Control of phase-matching factor $G$}

The phase-matching factor $G$ depends on the Gouy phase shift and dispersion in both the cadmium vapor and the argon buffer gas. These are experimentally controlled by adjusting the confocal parameter, oven temperature, and argon pressure, respectively.
In our setup, with $b = 6$ mm and 70 mbar argon pressure, the VUV yield is maximized when the oven controller reads 580~$^\circ$C, in reasonable agreement with the predicted optimum of 525 $^\circ$C~\cite{arxivpaper}. The residual discrepancy likely stems from uncertainties in temperature calibration and confocal parameter measurement, and spatial inhomogeneities in cadmium vapor density.

\noindent \textbf{Phase-matching near third-photon resonance}

Near third-photon resonance, $\chi^{(3)}$ is strongly enhanced, but phase matching is disrupted due to large VUV dispersion~\cite{arxivpaper}.
As $\lambda_3$ approaches resonance, the wavevector mismatch becomes too large for compensation by the Gouy phase shift, causing $G$ to vanish and suppressing VUV generation, as observed near 152.8 nm and 146.97 nm (see Fig. \ref{VUVoutput}d).
Phase-matching can be restored by tuning $\lambda_1$ near the $5\: ^1P_1$ or 5$^3P_1$ transitions and adjusting $\lambda_2$ accordingly to maintain two-photon resonance. The enhanced dispersion at $\lambda_1$ counteracts the VUV index change, restoring phasing matching. The triple resonance configuration dramatically boosts $\chi^{(3)}$ and thus the VUV output~\cite{CW2012}.

\noindent \textbf{Phase noise extraction from interference patterns}

To extract the relative phase between the two VUV beams, we analyse the displacement of interference fringes along the direction orthogonal to the fringe pattern. A 1D interference profile is obtained from each image along a calculated global sampling line, and the relative phase $\phi_i$ is extracted using a standard phase retrieval algorithm~\cite{fienup1982phase}.

The extracted phase $\phi_i$ is initially wrapped within the range $(-\pi, \pi]$. Over the full acquisition period, the phase may drift slowly by multiple $2\pi$ cycles. To reconstruct the true phase evolution, we apply phase unwrapping:
\begin{equation}
\Phi_i = \phi_i + 2\pi m_i,
\end{equation}
where $m_i$ is the integer number of $2\pi$ phase slips accumulated by the $i$-th image. We identify $m_i$ by imposing phase continuity between adjacent images:
\begin{equation}
\left| \Phi_i - \Phi_{i-1} \right| < \pi.
\end{equation}

The Allan deviation of the unwrapped phase $\sigma_\Phi(M,T_0)$ is calculated~\cite{riley1994allan}, and subsequently transformed~\cite{riley1994allan} into the Allan deviation of fractional frequency change using:
\begin{equation}
    \sigma_{y}(\tau) = \sigma_{y}(M,T_0) = \frac{\sigma_\Phi(M, T_0)}{2\pi M T_0 f_0},
\end{equation}
where $y = \Delta f / f_0$ is the fractional frequency change with $f_0$ the VUV frequency, $T_0 = 0.5$ s is the sampling interval, $M$ is the number of frames average, and $\tau = M T_0$ is the averaging time.

\noindent \textbf{High-frequency phase noise assessment}

The self-interference visibility $V_s(\tau)$ of a Lorentzian laser is~\cite{Mandel_Wolf_1995}:
\begin{equation}
V_s(\tau) = \frac{2\sqrt{I_1 I_2}}{I_1 + I_2} \mathrm{e}^{-\pi \nu_0 \tau},
\end{equation}
where $\tau$ is the delay time, $I_1$ and $I_2$ are the intensities of the interfering beams, and $\nu_0$ is the full width at half maximum (FWHM) of the laser linewidth. The delay time $\tau$ can also be interpreted as the effective observation time in a homodyne beat measurement against an ideal monochromatic reference, as both describe phase decoherence accumulated over time~\cite{Elliott1982}.

Assuming both VUV beams are independently and equally broadened with a Lorentzian profile, the visibility function is modified as
\begin{equation}
V_m(\tau) = \frac{2\sqrt{I_1 I_2}}{I_1 + I_2} \mathrm{e}^{-2\pi \nu_0 \tau},
\end{equation}
where the factor of 2 in the exponent accounts for the independent phase noise contributions from the two VUV beams.
Based on the measured fringe visibility of 0.5(1) with 1 s exposure, an upper bound of 0.08(2) Hz for the FWHM linewidth of a single VUV beam is obtained.

\noindent \textbf{Linewidth scaling under frequency multiplication}

The VUV phase noise $\Delta\phi_{\text{VUV}}$ is derived from phase fluctuations in the fundamental lasers~\cite{VUVCoherence2014,PeikVUV,1978phasenoise}, and follows
\begin{equation}
    \begin{aligned}             
    \Delta\phi_{\text{VUV}} 
    &= 2\Delta\phi_{375} +\Delta\phi_{710} \\
    &= 4\Delta\phi_{750} + \Delta\phi_{710},
    \end{aligned}
\end{equation}
where $\Delta\phi_{375}$ and $\Delta\phi_{710}$, $\Delta\phi_{750}$ represent the phase noise of the 375~nm, 710~nm, and 750~nm lasers, respectively.

Assuming white frequency noise dominates, the VUV linewidth, $\Delta f_{\text{VUV}}$, approximately scales with the time-averaged square of the phase fluctuations~\cite{DiDomenico:10}
\begin{equation}
    \begin{aligned}
    \Delta f_{\text{VUV}} & \approx  a \langle\Delta \phi_{\text{VUV}}^2\rangle \\
    & \approx 16\Delta f_{750} + \Delta f_{710} + 8a\langle\Delta \phi_{\text{750}}\Delta\phi_{\text{710}}\rangle,  \\
    \end{aligned}
\end{equation}
where $a$ is a proportionality constant, and the last term accounts for correlation.
In the limit of fully correlated phase noise, this cross term evaluates to $8a\langle\Delta \phi_{\text{750}}\Delta\phi_{\text{710}}\rangle = 8\sqrt{\Delta f_{750} \Delta f_{710}}$, yielding an upper bound for the VUV linewidth $\Delta f_{\text{VUV}}=50$ Hz. For uncorrelated lasers, the cross term vanishes, and the VUV linewidth reduces to $\Delta f_{\text{VUV}}=34$~Hz.

\clearpage

\end{document}